\documentstyle[multicol,aps,prl,epsfig]{revtex}
\tighten       %Gives single-space (RevTex 3.0)

\begin{document} 
\draft 
\title{Canted phase in double quantum dots}
\author{David S\'anchez, L. Brey, and Gloria Platero}
\address{Instituto de Ciencia de Materiales de Madrid (CSIC), 
Cantoblanco, 28049 Madrid, Spain.}
%\author{David S\'anchez}
%\email{dsanchez@icmm.csic.es}
%\affiliation{Instituto de Ciencia de Materiales de Madrid (CSIC), 
%Cantoblanco, 28049 Madrid, Spain.}
%\author{Luis Brey}
%\affiliation{Instituto de Ciencia de Materiales de Madrid (CSIC), 
%Cantoblanco, 28049 Madrid, Spain.}
%\author{Gloria Platero}
%\affiliation{Instituto de Ciencia de Materiales de Madrid (CSIC), 
%Cantoblanco, 28049 Madrid, Spain.}
\date{\today}
\maketitle
\begin{abstract}
We perform a Hartree-Fock calculation in order to
describe the ground state of a vertical double quantum dot
in the absence of magnetic fields parallel to the growth direction.
Intra- and interdot exchange interactions determine
the singlet or triplet character of the system as the tunneling is tuned.
At finite Zeeman splittings due to in-plane magnetic fields, we observe
the continuous quantum phase
transition from ferromagnetic to symmetric phase through a canted antiferromagnetic state.
The latter is obtained even at zero Zeeman energy for an odd electron number.
\end{abstract}
\pacs{73.21.La,73.43.-f,75.75.+a}
% 73.21.La=Electronic states in quantum dots
% 73.43.-f=Quantum Hall effects
% 75.75.+a=Magnetic properties of nanostructures
%\maketitle   

\begin{multicols}{2}
\narrowtext
\setcounter{equation}{0}
Interaction in two-dimensional (2D) electron gases leads to new quantum phases when more
degrees of freedom (external fields, spin and layer indices) are supplied to the system.
In bilayer $\nu=2$ quantum Hall (QH) structures theoretical calculations have predicted~\cite{das97}
and experimental evidence has confirmed~\cite{pel98}
the existence of a particularly exotic canted antiferromagnetic (C) phase
which continuously connect the naively-expected
ferromagnetic (F) and paramagnetic (P) ground states (GSs) as the layer separation is tuned.
Interesting predictions regarding C states in few-electron double quantum dots (DQDs)
in the QH regime have also been reported~\cite{mar00}.
Thus high magnetic fields seem an unavoidable condition
to observe this quantum transition
given that for vanishingly small magnetic fields F and C states in bilayer systems
would cost a good deal of kinetic energy.
Our goal is to challenge this idea by lowering the system dimensionality and
benefitting from the atomic-like spectrum of a semiconductor quantum dot~\cite{tar96}.
In DQDs (termed artificial molecules as well)
Coulomb blockade effects~\cite{liv96}, magnetization~\cite{oos98},
and the formation of a delocalized molecular GS~\cite{bli98} are some
of the exciting features observed.
%More recently,
%the discovery of new principles of computation based upon quantum mechanics
%has led to the idea of using coupled quantum dots for quantum computation~\cite{los98}.
%The recent fabrication of \emph{vertically} coupled quantum dots
%has triggered the theoretical interest in studying the physical properties of DQDs.
From the theoretical viewpoint 
exact diagonalization methods~\cite{ima99}, Hubbard-based models~\cite{ron99},
and spin density functional theories~\cite{par00} have been developed
to show the presence of magic-number, molecule-type, and Hund's-rule-violating states
in vertically coupled dots.

Here we present a Hartree-Fock (HF) theory for addressing many-body effects
in two vertically-coupled parabolic quantum dots
separated by a distance $a$ with a total electron number $N$.
We study this system in the absence of magnetic fields perpendicular to the dots.
Still, in order to add spin symmetry breakings
we allow for a parallel magnetic field
whose coupling to the electronic orbital motion is neglected
($a$ is assumed to be much smaller than the corresponding magnetic length).
We are interested in quantum dots whose atomic-like character results in
half-filled shells formed by quasi-degenerate eigenstates,
thus having large spin expected values acting as effective magnets.
Our main findings are: (i) the existence of a robust C phase
(envisaged as tilted spin vectors) at finite Zeeman energies for even values of $N$
linking the F (fully spin-polarized or, equivalently, triplet)
and P (fully isospin-polarized or singlet) GSs via a second-order phase transition;
(ii) the persistence of C states for $N$ odd even in the \emph{absence} of Zeeman gaps;
and (iii) the overture of a simple model which qualitatively explains our results,
allowing to deal with more complex quantum dot systems.

The electron spatial coordinates are denoted by $\vec r=(x,\vec\rho)$
where $\vec\rho=(y,z)$ and $x$ is the growth direction.
The wave function of the $i$-th electron may be expanded
in terms of 2D harmonic oscillator eigenstates, $\phi_{n l}(\vec \rho)$,
$n=0,1,2,\ldots$ being the radial quantum number and $l$ the angular momentum obeying
$l=-n,-n+2,\ldots,n-2,n$.
Accordingly, the Hamiltonian reads:
%\begin{widetext}
%\begin{eqnarray}
%{\cal H}  & = & \sum_{n_i l_i \sigma_i \alpha_i} \varepsilon_{n_i}
%c_{n_i l_i \sigma_i \alpha_i}^{\dag} c_{n_i l_i \sigma_i \alpha_i}
% + \sum_{n_i n_j n_k n_m l_i l_j l_k l_m \sigma_i \sigma_j \alpha_i \alpha_j \alpha_k \alpha_m}
%%{\cal V}_{n_i l_i \alpha_i, n_j l_j \alpha_j, n_k l_k \alpha_k, n_m l_m \alpha_m}
%{\cal V}_{n_i l_i \alpha_i, n_j l_j \alpha_j, n_k l_k \alpha_k, n_m l_m \alpha_m} \nonumber \\
%& \times & c_{n_i l_i \sigma_i \alpha_i}^{\dag} c_{n_j l_j \sigma_j \alpha_j}^{\dag}
%c_{n_k l_k \sigma_j \alpha_k} c_{n_m l_m \sigma_i \alpha_m} 
% - \sum_{n_i l_i \sigma_i \alpha_i \alpha_j} t \left[
%c_{n_i l_i \sigma_i \alpha_i}^{\dag} c_{n_i l_i \sigma_i \alpha_j}
%\left( 1-\delta_{\alpha_i \alpha_j} \right) + \mathrm{h.c.} \right] \, ,
%\label{eq-ham}
%\end{eqnarray}
%\end{widetext}
%\begin{eqnarray}
%{\cal H} & = & \sum \varepsilon_{n_i}
%c_{n_i l_i \sigma_i \alpha_i}^{\dag} c_{n_i l_i \sigma_i \alpha_i} 
%+ \frac{1}{2} \sum
%{\cal V}_{n_i l_i \alpha_i, n_j l_j \alpha_j, n_k l_k \alpha_k, n_m l_m \alpha_m} \nonumber \\
%& & \times c_{n_i l_i \sigma_i \alpha_i}^{\dag} c_{n_j l_j \sigma_j \alpha_j}^{\dag}
%c_{n_k l_k \sigma_j \alpha_k} c_{n_m l_m \sigma_i \alpha_m} \nonumber \\
%& & -\sum t \left[
%c_{n_i l_i \sigma_i \alpha_i}^{\dag} c_{n_i l_i \sigma_i \alpha_j}
%\left( 1-\delta_{\alpha_i \alpha_j} \right) + \mathrm{h.c.} \right] - \Delta_Z S_z\, ,
%\label{eq-ham}
%\end{eqnarray}
%\begin{eqnarray}
\begin{eqnarray}
{\cal H} & = & \sum \varepsilon_{n_i}
c_{n_i l_i \sigma_i \alpha_i}^{\dag} c_{n_i l_i \sigma_i \alpha_i} - 2 t I_x - \Delta_Z S_z  \nonumber \\
& + & \frac{1}{2} \sum
{\cal V}_{n_i l_i \alpha_i, n_j l_j \alpha_j, n_k l_k \alpha_k, n_m l_m \alpha_m} \nonumber \\
& \times & c_{n_i l_i \sigma_i \alpha_i}^{\dag} c_{n_j l_j \sigma_j \alpha_j}^{\dag}
c_{n_k l_k \sigma_j \alpha_k} c_{n_m l_m \sigma_i \alpha_m} \, ,
\label{eq-ham}
\end{eqnarray}
where the sums are extended to all indices,
$\sigma$ and $\alpha$ are the spin and layer indices, and
$\varepsilon_{n}=\hbar \omega (n+1)$ ($\hbar \omega$ is the confinement strength).
The second and third terms describe the coupling of the
isospin and the spin of the system with external perturbations,
namely the tunneling $t$ and the Zeeman splitting $\Delta_Z$.
%The $z$ ({\bf $x$??}) component of the isospin is defined as
%$I_z=\sum \left[ c_{n_i l_i \sigma_i \alpha_i}^{\dag} c_{n_i l_i \sigma_i \alpha_j}
%\left( 1-\delta_{\alpha_i \alpha_j} \right) + \mathrm{h.c.} \right]$.
The isospin points along $+$($-$)$z$ when the electron is at the top (bottom) layer.
Notice that the tunneling term only switches layer indices,
thus conserving the rest of quantum numbers.
$\Delta_Z=g\mu_B B$, where
$g$ is the Land\'e factor, $\mu_B$ the Bohr magneton,
and $B$ the applied magnetic field in the $z$ direction.
${\cal V}$ is the matrix element of the Coulomb potential
$V \left( \left| \vec r - \vec r\,' \right| \right)$.
%In the following $t>0$ is chosen so that bonding (antibonding)
%states have the lowest (highest) energy.

Because we seek to identify spin and particle-number broken symmetries which are reflected
in the interdot coherence,
we restrict our study to radial-symmetry-conserving solutions within an isolated dot
despite spontaneously radial symmetry breaking states might take place~\cite{con00}.
The resulting states $\psi$ can be labeled with \emph{individual} angular momenta:
$\psi_{l_i}(\vec r)=\sum_{n_i \sigma_i \alpha_i} d_{n_i \sigma_i \alpha_i} \,
\phi_{n_i l_i}(\vec\rho) \, f_{\alpha_i}(x) \, \chi_{\sigma_i}$.
$d$ are the coefficients of the expansion to be self-consistently calculated,
$f_{\alpha}(x)$ is the vertical component of the wave function,
and $\chi_{\sigma}$ is the spin part.
Hence the Hamiltonian is numerically diagonalized in separate $l$-subspaces
(though the matrix elements do depend on the total system configuration).

We consider that only the lowest-energy states in the vertical structure
are populated, and approximate the form of $f(x)$
as follows: $f_{\alpha={\rm T}} (x)=\sqrt{\delta(x)}$
($f_{\alpha={\rm B}} (x)=\sqrt{\delta(x-a)}$) for the top (bottom) layer.
More precise expressions for $f(x)$ would involve accurate form factors
entering the final results~\cite{cot92} but qualitatively all would remain the same.
Then taking a close look at the $x$ part of ${\cal V}$,
\begin{displaymath}
{\cal V} \propto \int dx \int dx' f_{\alpha_i}(x) f_{\alpha_j}(x')
V \left( \left| \vec r - \vec r\,' \right| \right) f_{\alpha_k}(x') f_{\alpha_m}(x) \,,
\end{displaymath}
we observe that the only terms different from zero are either
$\alpha_i=\alpha_j=\alpha_k=\alpha_m$ (\emph{intra}dot interaction) or
$\alpha_i=\alpha_m\neq\alpha_j=\alpha_k$ (\emph{inter}dot interaction)
since crossed terms (e.g., $\alpha_i=\alpha_k\neq\alpha_j=\alpha_m$) would be null
because top and bottom wave functions have zero overlap.

In the HF approach the electron-electron interaction part of the Hamiltonian
of Eq.~(\ref{eq-ham}) can be arranged in two parts:
the Hartree operator $\vartheta^H$,

%\begin{equation}
%{\cal V}^H=
%\sum_{n_i n_j n_k n_m l_i l_j l_k l_m \sigma_i \sigma_j \alpha_i \alpha_j \alpha_k \alpha_m}
%{\cal V}_{n_i l_i \alpha_i, n_j l_j \alpha_j, n_k l_k \alpha_k, n_m l_m \alpha_m}
%c_{n_i l_i \sigma_i \alpha_i}^{\dag} c_{n_m l_m \sigma_i \alpha_m}
%\left\langle c_{n_j l_j \sigma_j \alpha_j}^{\dag} c_{n_k l_k \sigma_j \alpha_k} \right\rangle
%\,,
%\end{equation}

%\begin{equation}
%{\cal V}^F=-
%\sum_{n_i n_j n_k n_m l_i l_j l_k l_m \sigma_i \sigma_j \alpha_i \alpha_j \alpha_k \alpha_m}
%{\cal V}_{n_i l_i \alpha_i, n_j l_j \alpha_j, n_k l_k \alpha_k, n_m l_m \alpha_m}
%c_{n_i l_i \sigma_i \alpha_i}^{\dag} c_{n_k l_k \sigma_j \alpha_k}
%\left\langle c_{n_j l_j \sigma_j \alpha_j}^{\dag} c_{n_m l_m \sigma_i \alpha_m} \right\rangle
%\,,
%\end{equation}

\begin{eqnarray}
\vartheta^H & = & \sum_{n_i n_m l_i l_m \sigma_i \alpha}
{\cal H}_{n_i n_m l_i l_m}
c_{n_i l_i \sigma_i \alpha}^{\dag} c_{n_m l_m \sigma_i \alpha}
\label{eq-har}
\end{eqnarray}
where
\begin{eqnarray}
{\cal H}_{n_i n_m l_i l_m} & \equiv & 
\sum_{n_j n_k l_j l_k \sigma_j \alpha'}
%{\cal V}_{n_i l_i \alpha_i, n_j l_j \alpha_j, n_k l_k \alpha_k, n_m l_m \alpha_m}
{\cal V}
\left\langle c_{n_j l_j \sigma_j \alpha'}^{\dag} c_{n_k l_k \sigma_j \alpha'} \right\rangle \,,
\label{eq-har2}
\end{eqnarray}
(the indices of ${\cal V}$ are omitted for the sake of simplifying notation)
and the exchange part $\vartheta^F$:
\begin{eqnarray}
\vartheta^F & = & -\sum_{n_i n_k l_i l_k \sigma_i \sigma_j \alpha \alpha'}
{\cal X}_{n_i n_k l_i l_k \sigma_i \sigma_j \alpha \alpha'}
c_{n_i l_i \sigma_i \alpha}^{\dag} c_{n_k l_k \sigma_j \alpha'}
\label{eq-exc}
\end{eqnarray}
where
\begin{eqnarray}
{\cal X}_{n_i n_k l_i l_k \sigma_i \sigma_j \alpha \alpha'} & \equiv & 
\sum_{n_j n_m l_j l_m}
%{\cal V}_{n_i l_i \alpha_i, n_j l_j \alpha_j, n_k l_k \alpha_k, n_m l_m \alpha_m}
{\cal V}
\left\langle c_{n_j l_j \sigma_j \alpha'}^{\dag} c_{n_m l_m \sigma_i \alpha} \right\rangle \,.
\label{eq-exc2}
\end{eqnarray}

Throughout our calculation we make
$\left\langle c_{l_i}^{\dag} c_{l_j}\right\rangle \propto \delta_{l_i,l_j}$
according to the aforementioned radial symmetry approximation.
In Eq.~(\ref{eq-har2}) $\alpha'=\alpha$ gives rise to intradot Hartree interaction.
The effect of this term is to make electrons repel from each other within the parabolic well.
The total energy of the dot is thus augmented.
Interdot Hartree interaction is naturally included for $\alpha'\neq\alpha$, notwithstanding
it does not have a strong influence in the final magnetic configurations, for it merely involves
a rigid shift of the energy levels. However, we keep it for having the same number
of electrons within each dot when $N$ is even and for obtaining
a more reliable value of the total energy of the system.
Intradot exchange interaction favors spin alignment within each dot as expected.
For $\alpha'\neq\alpha$ we are left with interdot exchange interaction.
We stress that although the barrier separating the double well in the $x$ direction
is wide enough and consequently the vertical parts of $\psi$ have negligible overlap
(in our case the overlap is strictly zero owing to the Dirac delta functions)
the interdot exchange interaction cannot be disregarded
as it plays a crucial role in the final DQD magnetic order.
In fact, it is the competition between the intradot exchange part
and the interdot exchange interaction
plus the tunneling term and the Zeeman energy
what drives the system from a GS dominated by intradot contributions
(large values of the interdot distance or small tunneling parameter)
to a state in which interdot effects prevail
(small values of $a$ or large $t$). In between non-trivial quantum phases can occur.
Our allowance of significant non-zero order parameters,
$\left\langle c_{\uparrow {\rm T}}^{\dag} c_{\downarrow {\rm B}}\right\rangle \neq 0$,
$\left\langle c_{\uparrow {\rm T}}^{\dag} c_{\downarrow {\rm T}}\right\rangle \neq 0$,
etc., eventually leads to spontaneous spin symmetry breakings,
spin rotations, canted phases and the fact that the particle number at each dot
is not a good quantum number.

At this point a small digression about the trustworthiness
of the HF model is needed.
Previous works have studied in detail~\cite{pfa93}
the differences between (un)restricted HF
theories and exact diagonalization methods in quantum dots.
It seems clear that a large number of electrons would result in a negligible
amount of quantum fluctuations which are unlikely to destroy mean-field-based predictions.
Furthermore, we can take advantage of large values of $\left\langle S_z \right\rangle$
(highest half-filled shells) to ease the appeareance of C phases.
Incidentally the existence of lower closed shells (hereafter designated as \emph{core})
is a crucial difference between QH systems
and DQDs in the absence of magnetic fields.
In the former case, only the lowest Landau level is occupied
and the kinetic energy plays a minor role.
In the latter, the dot fills its levels following an \emph{aufbau} rule,
thereby closing shells as $N$ is increased.
In principle we could fail to take into account the core states
and utilize them simply to assess the highest-lying eigenstates before coupling the dots
(then reducing the complexity of the calculation).
However, this would imply spin rotations applied solely to the last half-filled shell,
giving rise to a cost of energy which we are unable to evaluate \emph{a priori}
(as it is a self-consistent quantity).
The total number of electrons must then enter the numerical procedure.

Now, large values of $N$ tend to contract the
(renormalized) energy level interspacing in order to build a semiclassical radial density.
This may involve the drop of valence electrons below the core levels
and a subsequent reduction of $\left\langle S_z \right\rangle$.
A more favourable situation can be accomplished in part by enhancing the confinement.
Hence one should reach a compromise between these competing factors.

\begin{figure}
\centerline{
\epsfig{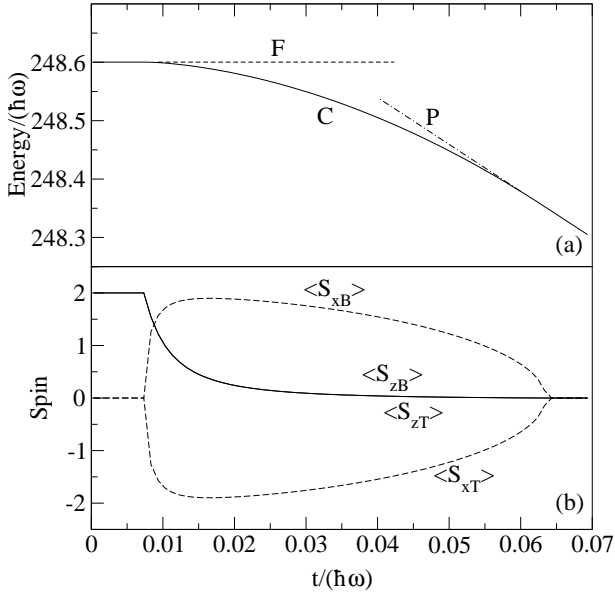}
}
%\begin{center}
%\includegraphics[angle=270,width=0.45\textwidth,clip]{fig1.eps}
%\end{center}
\caption{(a) Total energy of a DQD with $N=32$, $\hbar\omega=30$~meV, $a=l_{0}/2$,
and $\Delta_Z=0.017\hbar\omega$.
A dashed (dot-dashed) line shows the behaviour for the F (P) state
provided spontaneous spin symmetry breaking had not been taken place.
(b) Expectation values of the total spin for each dot.
$z$ ($x$) components are drawn in full (dashed) lines.
}
\label{fig1}
\end{figure}

The expansion of $\psi$ is enlarged enough, in such a way that the highest $n$ state
contributes less than 0.01\% to a typical density.
We present data for $N=32$ (though similar results are found for $N=18$),
setting $\hbar\omega=30$~meV and $a=l_{0}/2$, $l_{0}=\sqrt{\hbar / \left( m^{*}\omega \right)}$
being the harmonic oscillator typical length ($m^{*}$ is the GaAs effective mass).
As a result, the single-particle value of the $S_{z}$ projection onto the layer $\alpha$ is
$\left\langle S_{z\alpha} \right\rangle=2$.

\begin{figure}
\centerline{
\epsfig{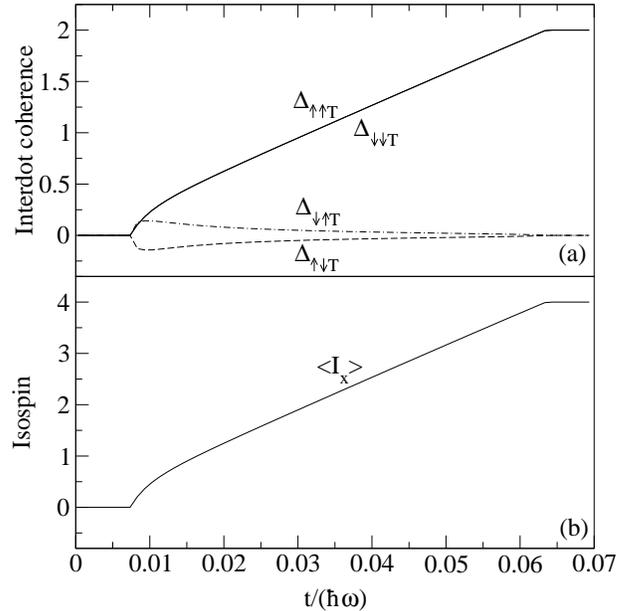}
}
%\begin{center}
%\includegraphics[angle=270,width=0.45\textwidth,clip]{fig2.eps}
%\end{center}
\caption{(a) Up-up and down-down spin interdot coherence (full lines) for the top layer.
Up-down (dashed line) and down-up (dot-dashed line) spin interdot coherence is also shown.
They are different from zero only in the C phase.
(b) $x$ component of the total isospin.
%$\left\langle I_{z} \right\rangle=0$
%throughout the tunneling range.
}
\label{fig2}
\end{figure}

Fig.~\ref{fig1}(a) depicts the total energy of the system,
$E=\left\langle{\cal H}\right\rangle$,
as a function of the tunneling parameter.
At low $t$ the GS is ferromagnetic, the intradot interaction being more important
than the interdot one plus the tunneling term.
Because this is a fully spin-polarized state, its energy does not depend on $t$
and remains constant until the system undergoes
a continuous quantum phase transition to the C phase.
In this case the system lowers its energy by increasing the tunneling contribution.
This favors the formation of singlets as well as the increase of interdot coherence
(see below).
To make the loss of Zeeman energy as
small as possible the spin configuration is then canted.
By further enhancement of $t$ the P phase is achieved
and a linear dependence of $E$ on the tunneling is obtained.
The C phase is thus a linear combination of the wave functions associated with F and P states.
In Fig.~\ref{fig1}(b) we have plotted $\left\langle S_{z\alpha} \right\rangle$ for both dots.
Its maximum is reached when the DQD is F, then it is depressed when entering the C phase.
The conversion to the singlet state ($\left\langle S_{z} \right\rangle=0$)
is continuous and $\left\langle S_{xT}\right\rangle=-\left\langle S_{xB} \right\rangle$.
This is the key feature of the appearance of a C phase--
total spin components in the plane perpendicular to the field
which are antiferromagnetically correlated.
These conclusions are reinforced by examining the interdot coherence of
the top (bottom) quantum dot:
$\Delta_{\sigma \sigma' {\rm T (B)}}=\sum_{n_i,l_i} \left\langle
c_{n_i l_i \sigma {\rm T (B)}}^{\dag}c_{n_i l_i \sigma' {\rm B (T)}} \right\rangle$.
It can be shown~\cite{bre98} that all the $\Delta_{\sigma \sigma' {\rm \alpha}}$ components
are zero in the triplet phase (see Fig.~\ref{fig2}(a)).
As $t$ increases, the system acquires interdot coherence until it is completely coherent
in the symmetric state
(for which $\Delta_{\uparrow \downarrow {\rm \alpha}}=\Delta_{\downarrow \uparrow {\rm \alpha}}=0$ and
$\Delta_{\uparrow \uparrow {\rm \alpha}}=\Delta_{\downarrow \downarrow {\rm \alpha}}\neq0$).
As C phases involve a spin symmetry breaking,
$\Delta_{\uparrow \downarrow {\rm \alpha}}=-\Delta_{\downarrow \uparrow {\rm \alpha}}$.
Moreover, since $\left\langle I_x\right\rangle$
yields half the difference between the number of electrons in symmetric states and
those in antisymmetric states,
$\left\langle I_{x} \right\rangle$ is zero in the F case and reaches its maximal
value in the P case.
As Fig.~\ref{fig2}(b) shows, the C phase develops intermediate
quantities.
The $z$ isospin component would be different from zero provided
there is more charge in one of the wells, e.g., by applying an electric bias; but
this is not the present case.

\begin{figure}
\centerline{
\epsfig{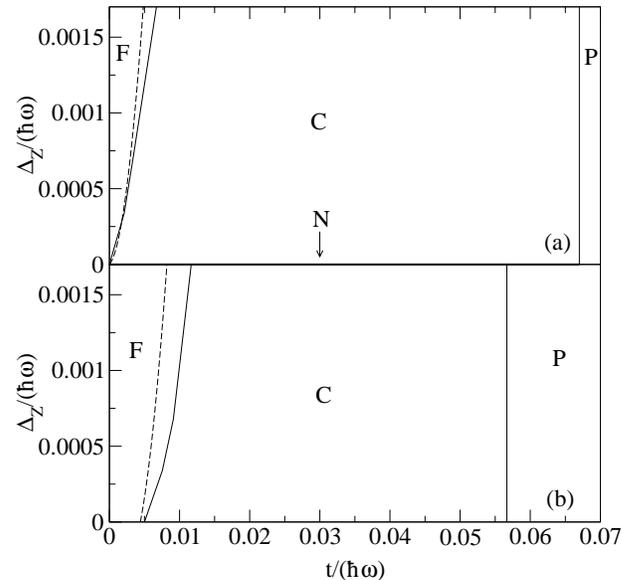}
}
%\begin{center}
%\includegraphics[angle=270,width=0.45\textwidth,clip]{fig3.eps}
%\end{center}
\caption{(a) Phase diagram for 32 electrons.
Full lines correspond to the numerical calculation.
The dashed line is obtained from a simple model.
The infinite slope of the C--P line seems to be correct
in the thermodynamic limit~\protect\cite{das97}.
(b) Same as (a) for $N$=31.
}
\label{fig3}
\end{figure}

In Fig.~\ref{fig3}(a) we draw the entire phase diagram which characterizes the distinct
GSs as a function of the Zeeman energy and the tunneling.
For large $\Delta_{Z}$ and small $t$ the DQD is in the spin-polarized
phase. In the opposite limit, the singlet state energy is lower. The C phase lies between them.
In the case of $\Delta_{Z}=0$ we obtain a purely antiferromagnetic or N\'eel (N) GS
with the spins pointing across from each other.

A more striking feature is observed when a hole is introduced into the system.
For $N$ odd the highest-lying shells are not closed and the remaining hole is
shared by the two dots.
From Fig.~\ref{fig3}(b) we see that the region covered by the C phase is reduced
at large $t$ because the system takes advantage more easily of the possibility of
tunneling by forming singlets.
Likewise the F state is more energetically favoured at low transmissions.
There is a range in the tunneling parameter at $\Delta_{Z}=0$
where the lack of charge spontaneously induces ferromagnetism.
But now the C state is extended even for $\Delta_{Z}=0$ since a N phase cannot
exist for an odd number of electrons.

A simple model may be aimed to shed light on this phenomena.
When the highest shell is occupied with an even number of electrons,
it is reasonable that a Heisenberg term accounts for the antiferromagnetic phase.
In addition, the total energy must include a contribution stemming from the Zeeman energy
which favours a parallel spin alignment.
Close to the F phase we propose the following energy functional:

\begin{eqnarray}
E_{t,\Delta_{Z}}(\theta) & = & -J(t) \vec S_{T} \cdot \vec S_{B}
- g\mu_{B} \vec B \cdot \left( \vec S_{T} + \vec S_{B}\right) \nonumber \\
& = & -J(t) S^2 \cos \theta - 2 \Delta_{Z} S \cos \left( \theta/2 \right) \,.
\label{eq-tJ}
\end{eqnarray}
Here we consider the total spins as classical
vectorial entities centered at each dot
and assume $\left| \vec S_{T}\right| =\left| \vec S_{B}\right| \equiv S$,
$\theta$ being the angle spanned by both vectors.
$J$($<0$) is a parameter fitted from the dependence of the total energy on $t$ at $\Delta_{Z}=0$
(one curve analogous to Fig.~\ref{fig1}(a)).
As a result, $J(t)$ is roughly quadratic for small $t$.
Then the critical line marking the transition to the canted phase from the F state
is achieved by setting $E''(\theta=0)=0$ which yields: $\Delta_{Z}(t)=-4J(t)S/2$.
This gives the piece of parabola showed in Fig.~\ref{fig3}(a).
The curve crosses the abscissa axis at $t=0$
proving that no F state can exist with $\Delta_{Z}=0$ at finite $t$ and the DQD takes on a N phase.
When an electron is removed from the DQD the remaining hole prefers to keep its spin parallel
to the rest while hopping from dot to dot. Therefore the system is sensitive to the
particular spin orientation of the hole and Eq.~(\ref{eq-tJ}) must include a term
accounting for this fact: $E\rightarrow E-t \cos \left( \theta/2 \right)$
(this is as if we had done an unitary rotation and kept only the diagonal terms)
resulting in $\Delta_{Z}(t)=\frac{t-4J(t)S^2}{2S}$.
Unlike the $N=32$ case, for $N=31$
$\theta=0$ is a minimum of $E$ at small values of $t$ 
(in the interval of physical interest, i.e., $\left[0,\pi\right]$)
and the system remains spin-polarized (see Fig.~\ref{fig3}(b)).
Larger $t$ means that $\theta=0$ corresponds to a relative maximum of $E(\theta)$
and one minimum at $\theta \neq 0$ shows up,
fulfilling that the C phase is now the lowest energy state.
Despite the simplicity of the model, the curves agree remarkably well
with the self-consistent numerical solutions. 

In summary, our analysis of the GS of a vertical DQD based
upon a mean-field framework predicts the existence of a canted phase 
for intermediate tunneling and not too high Zeeman energies.
For a sufficiently high even electron number
(for which quantum correlation effects are not expected
to qualitatively alter the conclusions)
the C phase continuously connects F and P states as
the tunneling parameter is varied.
When a hole is created in half-filled shells
the kinetic energy of the remaining electron promotes the F phase
at small $t$ and the C phase arises even at \emph{zero} (arbitrarily small) Zeeman splitting.

Thanks are due to C. Tejedor for useful discussions.
This work was supported by the Spanish DGES through Grants Nos. PB96-0875 and PB96-0085
and by the European Union TMR contract FMRX-CT98-0180.

\end{multicols}
\end{document}